# Strong in-plane magnetic anisotropy (Co$_{0.15}$Fe$_{0.85}$)$_5$GeTe$_2$/graphene van der Waals heterostructure spin-valve at room temperature

Roselle Ngaloy[1], Bing Zhao[1], Soheil Ershadrad[2], Rahul Gupta[3], Masoumeh Davoudiniya[2], Lakhan Bainsla[4,1], Lars Sjöström[1], Anamul M. Hoque[1], Alexei Kalaboukhov[1], Peter Svedlindh[3], Biplab Sanyal[2], Saroj P. Dash[1,5*]

[1]*Department of Microtechnology and Nanoscience, Chalmers University of Technology, SE-41296, Göteborg, Sweden*
[2]*Department of Physics and Astronomy, Uppsala University, Box-516, 75120 Uppsala, Sweden*
[3]*Department of Materials Science and Engineering, Uppsala University, Box 35, SE-751 03 Uppsala, Sweden*
[4]*Department of Physics, Indian Institute of Technology Ropar, Roopnagar 140001, Punjab, India*
[5]*Graphene Center, Chalmers University of Technology, SE-41296, Göteborg, Sweden*

**Abstract**

Van der Waals (vdW) magnets are promising owing to their tunable magnetic properties with doping or alloy composition, where the strength of magnetic interactions, their symmetry, and magnetic anisotropy can be tuned according to the desired application. However, most of the vdW magnet based spintronic devices are so far limited to cryogenic temperatures with magnetic anisotropies favouring out-of-plane or canted orientation of the magnetization. Here, we report room-temperature lateral spin-valve devices with strong in-plane magnetic anisotropy of the vdW ferromagnet (Co$_{0.15}$Fe$_{0.85}$)$_5$GeTe$_2$ (CFGT) in heterostructures with graphene. Magnetization measurements reveal above room-temperature ferromagnetism in CFGT with a strong in-plane magnetic anisotropy. Density functional theory calculations show that the magnitude of the anisotropy depends on the Co concentration and is caused by the substitution of Co in the outermost Fe layer. Heterostructures consisting of CFGT nanolayers and graphene were used to experimentally realize basic building blocks for spin valve devices such as efficient spin injection and detection. The spin transport and Hanle spin precession measurements prove a strong in-plane and negative spin polarization at the interface with graphene, which is supported by the calculated spin-polarized density of states of CFGT. The in-plane magnetization of CFGT at room temperature proves its usefulness in graphene lateral spin-valve devices, thus opening further opportunities for spintronic technologies.

**Introduction**

Magnetism in van der Waals (vdW) materials offers an excellent platform for exploring fascinating spintronic and quantum science and technology[1,2]. VdW magnetic materials are particularly interesting due to their tunable magnetic properties, where the magnetic anisotropy and Curie temperature ($T_c$) can be tuned by alloying[3,4], doping[5], gating[6,7], proximity-induced coupling[8], pressure[9], and functionalization[10]. The integration of vdW magnets in heterostructures with graphene, semiconductors, topological materials, and superconductors, can also result in unique proximity-induced effects and strongly correlated electronic phenomena[11].

By utilizing vdW magnets, several proof-of-concept spintronic and topological quantum phenomena have been demonstrated, such as spin-valves[12], tunnel magnetoresistance[13], proximity magnetism[14,15], exchange-bias[16], skyrmions[17], and spin-orbit torque[18,19]. However, most of the demonstrated devices were limited by the low Curie temperature ($T_c$) of the vdW magnets, limiting their potential for applications. After the discovery of $Fe_3GeTe_2$ with a $T_c$ around ~200 K[20], $Fe_5GeTe_2$ has been shown to exhibit itinerant ferromagnetism with higher $T_c$[21,22]. Interestingly, $Fe_5GeTe_2$/graphene spin-valve devices have been shown to possess canted perpendicular magnetic anisotropy at room temperature[23]. Encouragingly, well beyond room-temperature magnetic order can be achieved in $Co_xFe_{1-x}GeTe_2$[4,24], with the magnetic properties being tunable by the substitution of Fe with Co atoms. However, spintronic devices using such high $T_c$ vdW magnetic materials with magnetic anisotropy favouring in-plane magnetization are so far lacking.

Here, we demonstrate beyond room-temperature ferromagnetism in the vdW ferromagnet $(Co_{0.15}Fe_{0.85})_5GeTe_2$ (CFGT). We fabricated vdW based heterostructure spin-valve devices consisting of CFGT and chemical vapor deposited (CVD) graphene, which is a suitable channel material for long-distance spin transport and spin logic operations[25,26]. Detailed spin-valve and Hanle spin precession measurements on CFGT/graphene hybrid devices demonstrate efficient spin injection, detection, transport, and precession functionalities. The observation of symmetric Hanle spin precession signals proves the in-plane magnetization of CFGT, whereas the opposite sign of the measured spin signals provides evidence of negative spin polarization at the CFGT/graphene interface. These findings are well supported by density functional theory calculations, showing composition dependent modification of the magnetic anisotropy and spin polarized density of states.

**Results and Discussion**

In the vdW ferromagnet CFGT, the magnetic anisotropy and Curie temperature ($T_c$) are strongly dependent on Co doping, as schematically illustrated in Fig. 1a. To understand the preferential doping site of Co atoms, we calculated the formation energy of Co in each sublattice (details in Supplementary Information and Table S1). Formation energies indicate that up to 20.1% doping concentrations, Co atoms prefer to substitute to the outermost Fe sublattices, i. e. FeU and FeD species (dark blue color). This is compatible with previous results, showing that Fe split sites are most prone to defects[21,27,28]. The magnetic anisotropy energy (MAE) of pristine and Co doped structures in monolayer and bulk forms are listed in Table S2. Note that in the absence of Co, pure $Fe_5GeTe_2$ (FGT) monolayer has an out-of-plane easy axis, with MAE = +18.7 µeV/atom.

With Co doping, the direction of the easy axis changes to in-plane. In the monolayer regime, the MAE values for Co concentrations of 6.7, 13.4 and 20.1% are -29.6, -79.1 and -149.2 µeV/atom, respectively. A similar trend is also observed in the bulk samples. A direct relation between the Co concentration and the MAE is evident, where doping can alter the magnetic anisotropy from a weak out-of-plane to a stronger in-plane anisotropy.

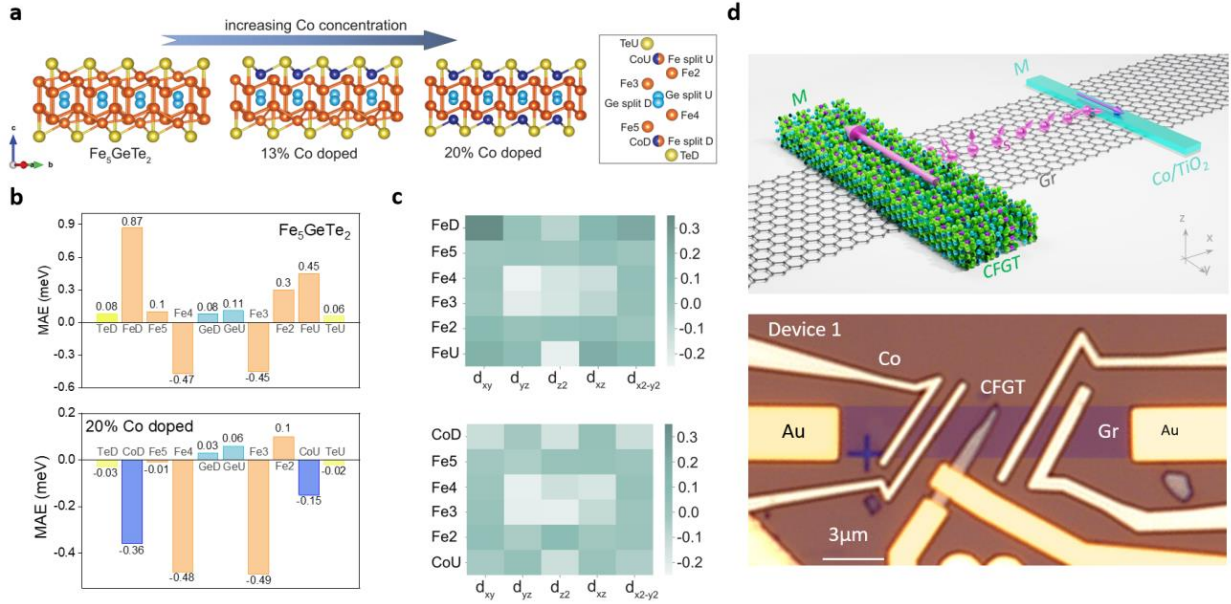

*Figure 1. Composition-dependent magnetic anisotropy calculations of CFGT and CFGT/graphene van der Waals spin-valve device. a.* Schematic representation of CFGT structure as the concentration of Co atoms increases, occupying the outermost (split) sublattice of Fe, where dark blue, orange, yellow, and light blue atoms represent Co, Fe, Te, and Ge, respectively. The right panel shows atomic notations and their relative positions along the c axis. *b.* Atom projected MAE in meV for pure FGT and 20.1% doped CFGT in upper and lower panels, respectively. *c.* Heat map of orbital projected MAE resolved into 3d orbitals of Fe and Co atoms for pure FGT and 20.1% doped CFGT in upper and lower panels, respectively. *d.* Schematics and microscope picture of CFGT/graphene lateral spin-valve device, where CFGT acts as a spin injector/detector and $Co/TiO_2$ as reference ferromagnetic contact on the CVD graphene channel. Non-magnetic Ti/Au contacts are used for reference electrodes. The scale bar is 3 µm.

To understand the origin of this magnetic anisotropy change, we have calculated the atom and orbital projected MAE for pure FGT and 20.1% CFGT, as shown in Fig. 1b and c, respectively (for intermediate concentrations see Supplementary Information Fig. S1). In the case of pure FGT (Fig. 1b upper panel), the main contribution to MAE comes from Fe atoms while Ge and Te atoms have a marginal impact on the direction of the easy axis. Among the Fe sublattices, however, the contribution is not uniform. Outermost Fe atoms have large out-of-plane easy axis (0.87 and 0.45 meV per FeD and FeU, respectively). As we move towards the center of a monolayer, the size of out-of-plane magnetic anisotropy on Fe5 and Fe2 sublattices decreases to 0.10 for Fe5 and 0.30 meV/atom for Fe2. Notably, in the central region, Fe3 and Fe2 sublattices have large in-plane MAE of -0.47 and -0.45 meV/atom, respectively. Therefore, there is a competing interplay between the direction of magnetic anisotropy of various Fe sublattices that determines the overall easy axis of the pristine FGT crystal. In 20.1% doped CFGT, shown in Fig. 1b (lower panel), Co atoms (in dark blue bars), which have substituted Fe split site atoms, have a relatively large in-plane magnetic easy axis with -0.30 for CoD and -0.15 meV/atom for CoU. Furthermore, we observe a decrease in the MAE values of the Fe5 and Fe2 sublattices, located next to Co dopants, with an MAE of -0.01 and 0.1 meV/atom, respectively. The central sublattices of Fe4 and Fe3 are marginally affected by Co atoms. Accordingly, one can understand that two congruent contributions add up to the change in the easy axis in the presence of Co atoms. Firstly, and most importantly, is the substitution of largely out-of-plane FeU and FeD with in-plane CoU and CoD. Secondly, is the reduction of 0.1 and 0.2 meV/atom on Fe5 and Fe2 atoms, respectively, which are located adjacent to Co dopants. The atom projected MAEs for lower Co concentrations also show similar behaviors (see Supplementary Information Fig. S1), indicating that the physics behind the easy axis change acts independent of configuration or doping concentration.

We further provided the heat map of orbital resolved MAE for d-orbitals of Fe and Co atoms in Fig. 1c. It was found that in pure FGT (upper panel), the $d_{xy}$ and $d_{x2-y2}$ orbitals contribute strongly to the out-of-plane anisotropy of FeU and FeD atoms, while the in-plane tendency in Fe3 and Fe4 mainly comes from the $d_{yz}$, $d_{xz}$, and $d_{z2}$ orbitals. In contrast, in the Co dopants (CoU and CoD) in CFGT (lower panel), the $d_{xy}$ and $d_{x2-y2}$ orbitals tend to have in-plane moments. Table S3 in the Supplementary Information shows the average magnetic moments on different sublattices in FGT and 20.1% doped CFGT. One can see that Co atoms have a magnetic moment of 0.58 and 0.80 $\mu_B$, depending on their occupational site. This is significantly smaller than that of Fe split atoms (with 1.53 and 1.93 $\mu_B$ for FeU and FeD, respectively). Thus, Co atoms prefer to have a lower spin state compared to Fe atoms in the same occupational site.

To take advantage of the tunable magnetic anisotropy and improved $T_C$, we fabricated lateral graphene spin-valve devices (Fig. 1d) using CFGT grown by chemical vapour transport (CVT) method (from HqGraphene). A lateral spin-valve is a basic building block of spintronic devices, where one can investigate several functionalities, such as spin injection, transport, precession and detection using novel vdW materials and hybrid structures. A schematic diagram and optical microscope picture of a lateral spin-valve device are illustrated in Fig. 1d, where the spin current can be injected/detected by CFGT on a CVD graphene channel and detected/injected by a reference ferromagnetic Co/TiO$_2$ contact (see Methods section for fabrication details). As the aspect ratio of CFGT is an important factor in determining the magnetic shape anisotropy and the easy axis of magnetization, a very narrow CFGT flake ($W_{CFGT}$≈0.7 µm and thickness of ~30 nm) was used to achieve a stronger magnetic shape anisotropy and a nearly single magnetic domain state at the interface with graphene. The well-known in-plane magnetic anisotropy and spin polarization of Co/TiO$_2$ electrodes on our graphene channels should assist us in quantifying the magnetic properties of new vdW magnets.

First, the temperature-dependent magnetic moment of a bulk CFGT crystal measured with a superconducting quantum interference device (SQUID) magnetometer shows that a higher $T_C$ above 300 K has been achieved (Fig. 2a). Magnetic hysteresis loops measured from 10 to 300 K for both in-plane (B//ab) and out-of-plane (B//c) orientations (Fig. 2b,c) of the magnetic field show that a strong in-plane magnetic anisotropy is maintained up to room temperature. The magnified hysteresis loop at 300 K (Fig. 2b) shows a clear remanence in the in-plane direction, whereas the magnetic hysteresis almost vanishes for the out-of-plane orientation. Such stabilization of in-plane magnetic anisotropy agrees with our theoretical calculations presented above.

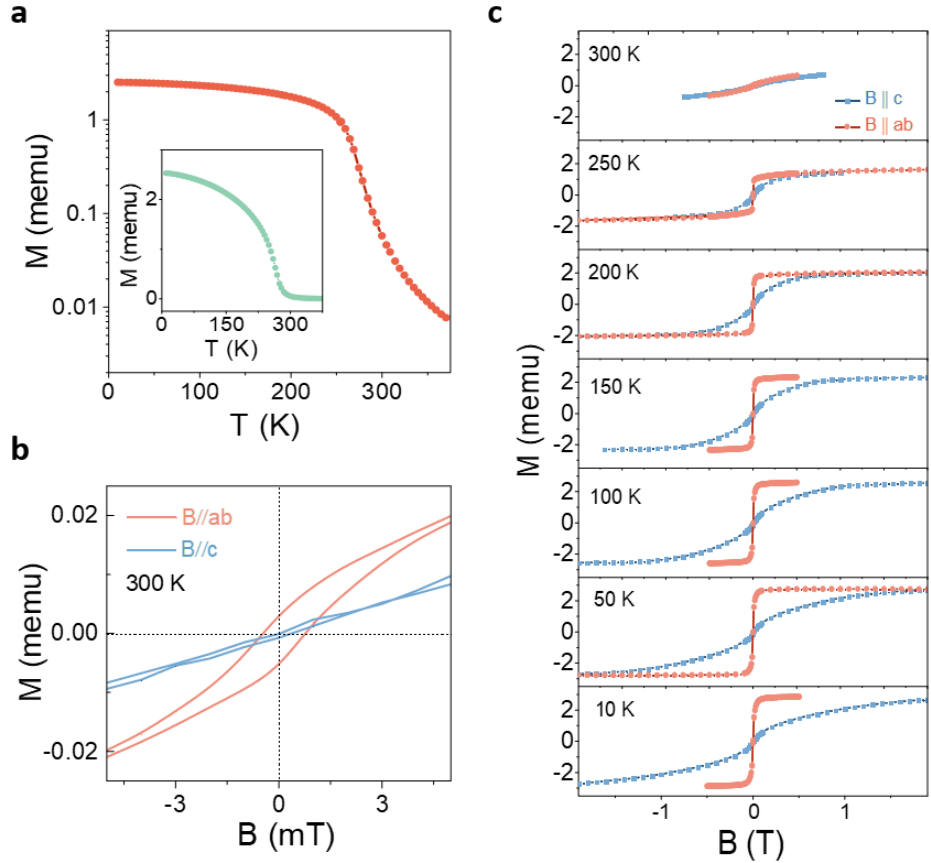

*Figure 2. Beyond room temperature magnetism in CFGT. a.* Magnetic moment (M) on a logarithmic scale of bulk CFGT versus temperature for an applied magnetic field of 50 mT. The inset shows the M(T) behavior of CFGT using a linear scale for the magnetic moment. *b.* Magnified magnetic hysteresis loops of CFGT with in-plane (orange) and out-of-plane (light blue) magnetic field sweeps at 300 K. *c.* Magnetic hysteresis loops of CFGT crystal with in-plane (orange) and out-of-plane (light blue) magnetic field sweeps in the temperature range of 10-300 K, showing strong in-plane magnetic anisotropy maintained through the temperature range.

The lateral CFGT/graphene heterostructure spin-valve device can probe the interfacial spin polarization and hence the magnetic anisotropy of the thin CFGT flakes. We performed non-local spin transport measurements on graphene spin valve to quantify the injected (detected) spin current by a thin CFGT flake (~ 30 nm). Figure 3a illustrates the non-local measurement configuration, where the current is passed through the Co-graphene interface (injector circuit), and voltage is measured at the CFGT-graphene interface (detector circuit). Figure 3b shows the nonlocal spin-valve data for the detection of spin current by CFGT (Dev 1) with the magnetic field sweep along the y-axis ($B_y$) at room temperature. This allows the control of the relative orientation of the magnetic moment of the injector (Co) and detector (CFGT) from parallel to anti-parallel orientations resulting in the spin-valve signal with two resistance states. Since the reference Co contacts have a magnetic easy axis along the y direction, the observation of a spin-valve signal confirms the detection of in-plane $S_y$ spins by the detector CFGT contact. Noticeably, the CFGT shows a sharp switching with a clear remanence and hysteresis in the spin-valve signals, indicating the presence of dominant $S_y$ spin polarization and in-plane magnetic anisotropy. This is also replicated in minor loop measurements with forward and backward field sweeps before reaching the Co coercive field. The data measured in two magnetization configurations show an apparent memory effect (Fig. 3c). Spin detection by CFGT using a different contact in Dev 1 is presented in Supplementary Information Fig. S2, and spin injection by CFGT is also measured in another device (Dev 2) as presented in Supplementary Information Fig. S3.

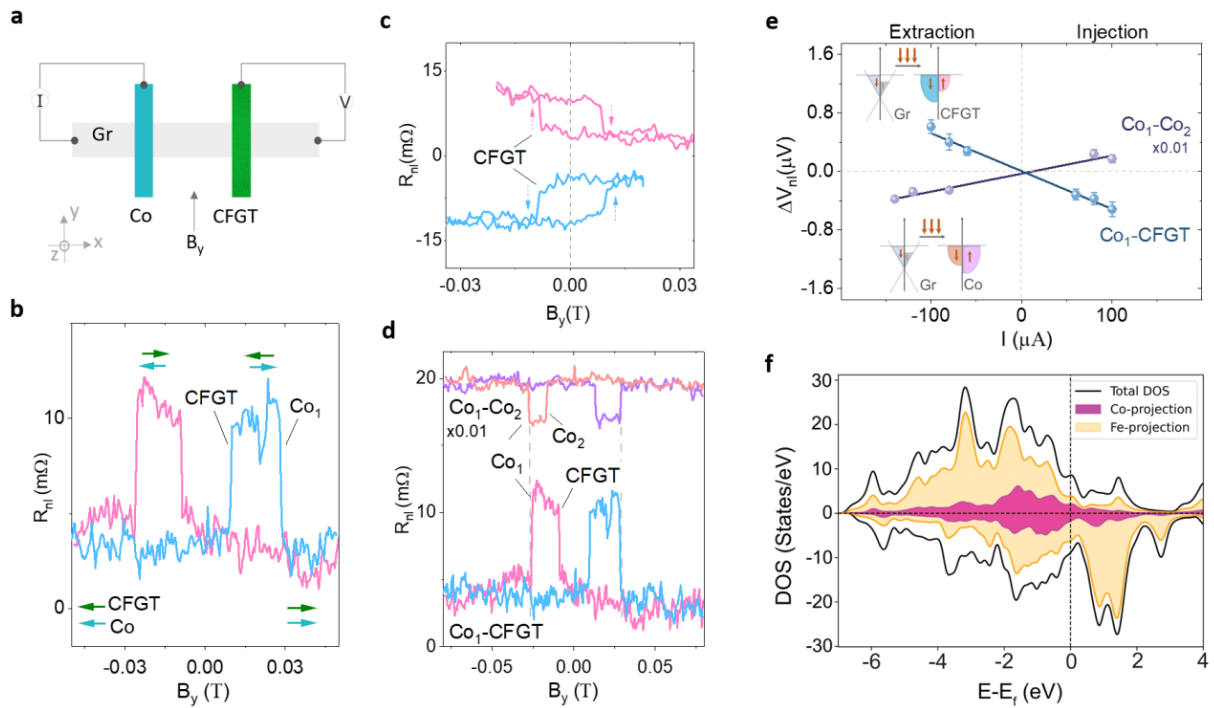

*Figure 3. Lateral spin-valve with CFGT-graphene heterostructure at room temperature. a. Schematic diagram for non-local spin-valve for Co(injection)-Gr-CFGT (detection) configuration for Dev 1. b. Measured spin-valve signal for Dev 1, showing sharp switching for both CFGT and Co magnetic contacts. The arrows represent the direction of magnetization of CFGT and Co through the magnetic sweep. c. Minor-loop measurements of CFGT. The arrows indicate the CFGT magnetic moment switching under the up- and down-sweep of magnetic field. d. Comparison of measured nonlocal spin-valve signals of Co-Gr-CFGT and Co-Gr-Co devices, showing the opposite sign for the same polarity of applied bias current -100 µA. The signals are shifted along the y-axis and the signal for Co-Gr-Co was rescaled by x0.01 for clarity. e. Bias dependence of spin signal for both Co-Gr-CFGT and Co-Gr-Co spin-valve devices in the spin injection and extraction regimes, showing opposite spin polarization for Co and CFGT contacts. The error bars are estimated from the standard deviation of the measured data. The insets show the detection mechanism of spin current for $Co_1$-CFGT and $Co_1$-$Co_2$ devices. f. DOS (solid black line) and pDOS (colored shaded) for 20.1% doped CFGT, where orange represents the density of Fe states and magenta represents the density of Co states.*

We performed control measurements to probe the sign of the spin polarization at the CFGT/graphene interface and compared it to the standard Co/graphene contacts by measuring a purely Co-Co (injector-detector) device. It is well established that Co has a positive spin polarization, which means that the majority of spins at the Fermi level are parallel to its bulk magnetization[29]. However, we observed a reversal of the spin-valve signal when we compare Co-CFGT to Co-Co at the same bias condition (Fig. 3d). We plot the detailed bias-dependent spin signal for Dev 1 in Fig. 3e, where the polarity of the spin-valve signal for the Co-CFGT and Co-Co devices change according to the polarity of the bias current and follows a linear trend for the range of current considered. Note that the opposite polarity for the measured spin-valve signals for Co-CFGT and Co-Co, remains consistent throughout the bias range considered. For different current bias polarities, the mechanism for spin accumulation at the interface of the injector magnet and the graphene transport channel changes from either spin injection (+$I_{dc}$) or extraction (-$I_{dc}$). Considering the Co injector, for positive bias, spin-polarized electrons tunnel from Co into the graphene channel, accumulating a spin population at the interface with polarization in accordance with the spin polarization of Co. On the other hand, when we apply a negative bias, electrons from graphene tunnel into the injector. In this case, since there are more available states for majority spin electrons in Co, more majority spin electrons will be extracted from the graphene channel, creating a non-equilibrium spin population in graphene dominated by minority spin electrons, as illustrated in the inset of Fig. 3e. This can further be analyzed by looking at the expression for the amplitude of the nonlocal spin-valve signal given below[30]:

$$\Delta R_{NL} = \frac{P_{CFGT} \cdot P_{Co} \cdot \lambda_{gr} \cdot R_{sq}}{2 w_{gr}} \cdot e^{(-L_{ch}/\lambda_{gr})} \qquad (1)$$

where $P_{CFGT}$ and $P_{Co}$ are the spin polarization of CFGT and Co, respectively; $\lambda_{gr}$ is the spin diffusion length in graphene, $R_{sq}$ is the graphene square resistance, $L_{ch}$ and $w_{gr}$ are the graphene channel length and width, respectively.

The opposite nonlocal spin-valve signals observed in these devices indicate opposite spin polarizations of the Co and CFGT contacts on graphene. This means that CFGT has a negative spin polarization and that the minority density of states is larger than the majority density of states at the Fermi level. In this case, when we apply a bias on the Co injector, and align the magnetic moments of Co and CFGT, the spin polarizations of the detector and injector have opposite directions, hence opposite signs are expected for the Co-Co and Co-CFGT devices in accordance with eq. 1. This observation is further substantiated by our calculations of the density of states (DOS) and projected density of states (pDOS) for CFGT. Figure 3f and Fig. S4 show the total DOS and pDOS for 20.1% and 13.4% doped CFGT, respectively. The spin polarization can be calculated by using the expression: $Spin\ polarization = \frac{N\uparrow - N\downarrow}{N\uparrow + N\downarrow}$, where N is the density of states at the Fermi level (solid vertical black line), and ↑ and ↓ correspond to the majority and minority spin directions, respectively. Accordingly, a negative spin polarization of -2.1% and -7.9% was calculated for 20.1% and 13.4% CFGT, respectively. These are smaller than that of pure FGT with -10% spin polarization[23]. The accumulated pDOS of the Fe atoms, shown by orange shade, has the largest contribution to the total DOS and thus, the two curves resemble each other significantly. However, right at the Fermi level, the Co atoms have a positive spin polarization (violet shaded curve), which competes with the overall negative polarization of the Fe atoms, decreasing the magnitude of the negative spin polarization. Consequently, the magnitude of spin polarization should decrease as the Co atom concentration increases, hence CFGT has a smaller spin polarization compared to pure FGT. Atom-resolved spin polarizations are listed in Supplementary Information Table S3. Here, the main compensation comes from CoU with +17.5% spin polarization, substituting FeU with -31.6% spin polarization.

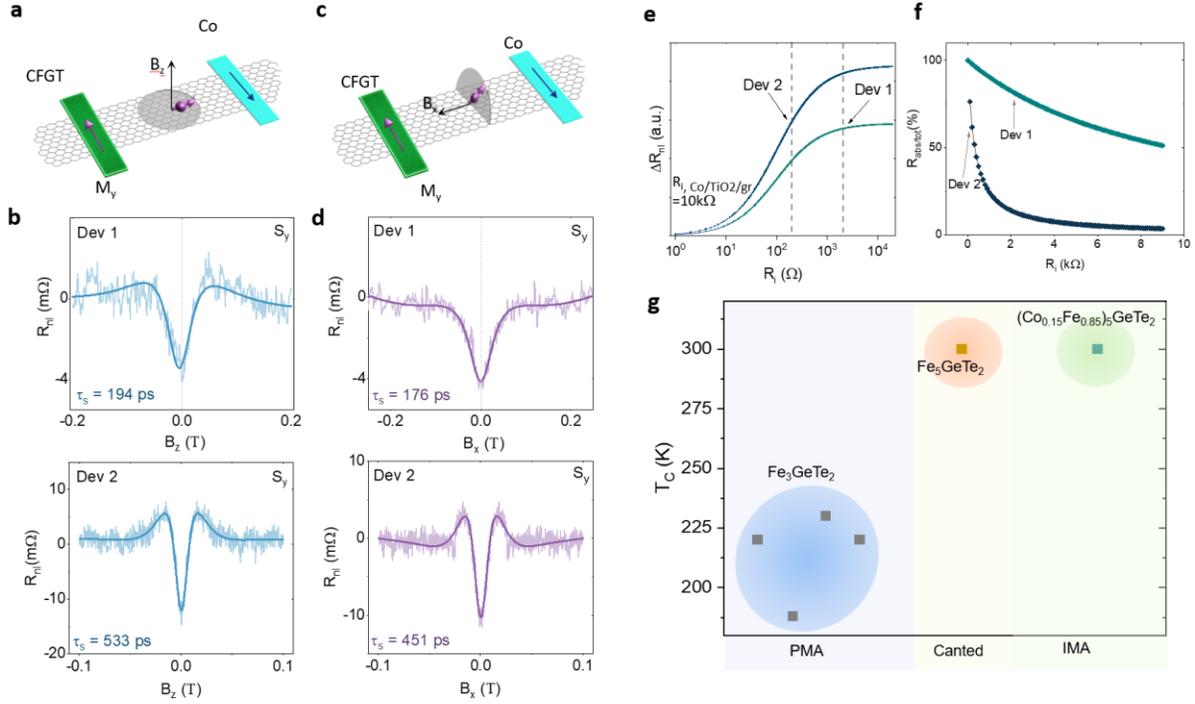

**Figure 4.** *Hanle spin precession measurements in CFGT-graphene heterostructure at room temperature*. **a.** *Schematic diagram of z-Hanle spin precession measurement in the CFGT-graphene spin-valve. The magnetic field is applied along the z-direction and $M_y$ is the magnetic moment of CFGT along the y-axis.* **b.** *Measured symmetric z-Hanle spin precession signal and the data fitting for Dev 1 (CFGT as detector) and Dev 2 (CFGT as injector), showing the dominant $S_y$ spin polarization component of CFGT at room temperature. A linear background is subtracted from the measured data.* **c.** *Schematic diagram of Hanle spin precession measurement showing applied magnetic field along x-direction. Similarly, $M_y$ is the magnetic moment of CFGT along the y-axis.* **d.** *The measured symmetric x-Hanle spin precession signal and data fitting for Dev 1 and Dev 2 show the dominant $S_y$ spin polarization component of CFGT at room temperature. A linear background is subtracted from the measured data.* **e.** *Calculated nonlocal spin signal as a function of the CFGT-graphene interface resistance $R_i$ in the CFGT/Gr/Co lateral spin-valve device. Dashed lines indicate the measured CFGT-Gr interface resistance in our devices.* **f.** *Calculated spin absorption ratio as a function of the interface resistance $R_i$ and square resistance $R_{sq}$ of the graphene channel.* **g.** *Comparison of our results to reported magnetic tunnel junctions and spin-valve devices based on the family of vdW ferromagnets $Fe_3GeTe_2$ and $Fe_5GeTe_2$*[12,13,23,31,32].

To unambiguously prove the spin signal in our devices, we conducted Hanle spin precession experiments[33] in both the z-axis (zHanle) and the x-axis (xHanle) geometries, which helps to estimate the tunnel spin polarization of CFGT/graphene interface in different orientations and evaluate the spin lifetime and diffusion length in the graphene channel. In zHanle, the out-of-plane magnetic field $B_z$ drives the injected spins to precess in the x-y plane, as illustrated in Fig. 4a. Spin precession results in a $B_z$-dependent evolution of the nonlocal spin signal. The measured nonlocal Hanle signal is proportional to the spin polarization of injector and detector ($R_{nl} \propto P_{in} \cdot P_{de}$), with $P_{in(de)}$ being the spin polarization of the injector (detector), and a symmetric (antisymmetric) signal corresponds to an initial $S_{y(x)}$ spin state. As presented in Fig. 4b, for Dev 1, the data for the nonlocal zHanle signal is a symmetric curve suggesting a parallel state of the injected spins $S_y$ with the magnetic moments of the CFGT detector. For Dev 2, the signal is obtained by pre-setting the injector (CFGT) and detector (Co) to parallel and anti-parallel states (the raw data and analysis are shown in Fig. S5). The decomposed average zHanle signal only shows symmetric components, which means that, like Dev 1, the injected spins by CFGT only have components along the y-axis, $S_y$.

By fitting the measured signal using the Hanle formula (eq. 2)[30], we evaluate the spin lifetime and the spin diffusion length in graphene channel to be $\tau_s = 194 \pm 29\ ps$ and $\lambda_{gr} = 2.1 \pm 0.3\ \mu m$ for Dev 1 and $\tau_s = 533 \pm 20\ ps$ and $\lambda_{gr} = 4.27 \pm 0.2\ \mu m$ for Dev 2, respectively.

$$R_{NL} \sim \int_0^\infty \sqrt{\frac{1}{4\pi t}} \exp\left[\frac{-L_{ch}^2}{4D_s t}\right] \cos(\omega t) \exp\left[-\frac{t}{\tau_s}\right] dt \qquad (2)$$

where $L_{ch}$ is the graphene channel length, $D_s$ is the spin diffusion constant, $\tau_s$ is the spin lifetime, and $\omega$ is the Larmor precession frequency.

Similarly, the x-Hanle measurement is performed with the external $B_x$ field applied along the x-axis, inducing a spin precession in the yz-plane (Fig. 4c). The observation of a symmetric x-Hanle curve in Fig. 4d, for Dev 1 and Dev 2, demonstrates the dominant in-plane spin component $S_y$ at the CFGT/graphene interface. From the Hanle fitting, we extract the spin parameters $\tau_s = 176 \pm 14\ ps$ and $\lambda_s = 2.5 \pm 0.2\ \mu m$ for Dev 1, and $\tau_s = 451 \pm 18\ ps$ and $\lambda_s = 3.42 \pm 0.1\ \mu m$ for Dev 2 in the graphene channel. Both sets of data are comparable to the spin transport parameters obtained from the z-Hanle data.

We can extract the spin polarization at the CFGT/Gr interface (for Dev 2) to be 4.93% and 4.5% using eqs. 1 and 2, respectively, considering a measured spin polarization of 5.8% in the reference Co/TiO$_2$/Gr contact (spin transport in the Co-Co reference for Dev 2 is presented in Supplementary Fig. S6). The spin injection efficiency can be influenced by the conductance mismatch and spin absorption effects between the ferromagnet and graphene. Since the resistance in graphene and ferromagnets differ by a few orders of magnitude, usually a tunnel barrier is introduced to improve conductance matching[34]. To assess the impact of the conductance mismatch between CFGT and graphene we look at the effect of the CFGT/Gr interface resistance on the nonlocal spin-valve signal by employing the drift-diffusion model[35]. For our calculations shown in Fig. 4e, we set the Co/TiO$_2$/Gr interface resistance to 10 kΩ, which is in the same range as the values measured for Dev 1 and Dev 2. The difference between calculated and nominal values for the two devices stems from the different $R_{sq}$ of the graphene channel (Fig. S7). For Dev 1, the CFGT/Gr interface resistance is 2 kΩ, which is sufficient to obtain near-optimal nonlocal signal, for this specific set of parameters. For Dev 2, with a more transparent contact of 200 Ω, the calculated signal obtained is lower than the nominal nonlocal signal, reflecting a more pronounced effect of conductance mismatch in Dev 2. With optimal conductance matching, the projected spin polarization could reach up to ~6.57%, which is in good agreement with the theoretical calculations discussed above.

To evaluate how the CFGT affects spin absorption in the graphene transport channel, we calculate the spin absorption rate $\Gamma$ using an approximate model:[36]

$$\Gamma = \frac{R_{sq}D_s}{\rho_{CFGT}\lambda_{CFGT} + R_i A} \quad (3)$$

where $\rho_{CFGT}$ and $\lambda_{CFGT}$ are the spin resistivity and spin diffusion length in CFGT, respectively. $R_i$ is the interface resistance of the CFGT/Gr heterostructure and $A$ refers to the interfacial area. Due to CFGT's magnetic nature $\rho_{CFGT} \cdot \lambda_{CFGT}$ becomes significantly less[37] than $R_i \cdot A$, resulting in a spin absorption rate that is dominated by the graphene resistance, spin relaxation in graphene, and interface resistance. Furthermore, the total spin relaxation time can be expressed as $\tau_{s,total} = \tau_{soc} + \tau_{abs}$, where $\tau_{abs} = 1/\Gamma$ and $\tau_{SOC}$, spin lifetime from SOC, was extracted from the spin parameters obtained in the experiment [38]. We obtained a spin absorption contribution of 80% for Dev 1, and 61% for Dev 2 as illustrated in Fig. 4f. The higher spin absorption for Dev 1, is in part due to the high graphene channel resistance of ~3.3 kΩ which is one magnitude higher compared to Dev 2 (Fig. S7). By increasing the interface resistance, which could be done by integrating a tunnel barrier, we can reduce the spin absorption contribution.

From both theoretical and experimental findings, we can conclude the presence of a strong in-plane magnetic anisotropy in CFGT and in-plane spin polarization at the CFGT/graphene interface at room temperature. In comparison to earlier reports for the family of FGT vdW magnets (Fig. 4g), our results highlight the unique in-plane anisotropy and high $T_C$ of CFGT in contrast to the perpendicular magnetic anisotropy (PMA) in $Fe_3GeTe_2$ and canted magnetization in $Fe_5GeTe_2$.

**Summary**

In summary, our findings reveal room-temperature ferromagnetism in the vdW magnetic metal CFGT with a strong in-plane magnetic anisotropy. Based on DFT calculations, this strong in-plane magnetic anisotropy stems from the substitution of the outermost Fe atoms, having an out-of-plane easy axis, with Co dopants, having an in-plane easy axis. We demonstrated the utilization of such a high $T_C$ vdW magnet in CFGT/graphene heterostructure spin-valve devices at room temperature. The in-plane magnetic anisotropy and spin polarization in CFGT were probed using spin-valve and Hanle spin precession measurement geometries, which provide unique insights into the room temperature magnetism and spin polarization at the CFGT-graphene heterostructure interface. Spin injection, detection and transport have been observed with a negative spin polarization of ~ 5% at the CFGT/graphene interface at room temperature. The symmetric Hanle curves measured in the devices prove the in-plane spin polarization of the CFGT at room temperature, which was further verified through DFT calculations. These results establish that integration of in-plane vdW magnets with graphene in spin-valve devices has considerable potential in the development of room temperature 2D spintronic applications.

**Methods**

**DFT Calculations:** Structural optimization and formation energy calculations were done by the Vienna Ab initio Simulation Package (VASP)[39,40]. The exchange-correlation potential was approximated by the generalized gradient approximation (GGA) with the Perdew, Burke, and Ernzerhof (PBE) functional[41]. For integration in the Brillouin zone, we used a 11×11×1 and 11×11×3 k-point grid in the Monkhorst-Pack scheme[42] for a √3 × √3 supercell of monolayer and bulk CFGT, respectively. The equilibrium lattice constants and atomic positions were optimized through energy minimization, using the conjugate gradient method up to the point that the force components on each atom were below 0.01 eV/Å. In the monolayer regime, the interaction between periodic images along the z-axis was minimized by adding a vacuum spacing of at least 20 Å. In all calculations, vdW correction via DFT-D3 method of Grimme with zero-damping function was enabled. The electronic and magnetic properties were calculated by the QuantumATK-Synopsys package[43,44], using LCAO basis set, and "PseudoDojo" presudopotential[45]. A density mesh cut-off of 140 Hartree and a k-point grid of 15×15×1 and 15×15×3 were used for monolayer and bulk self-consistent calculations. The magnetic anisotropy energy was calculated based on the force theorem, with a k-point grid of 25×25×1 and 25×25×3 for monolayer and bulk, respectively, using the expression: $MAE = E\perp - E\parallel$, where $E\perp$ and $E\parallel$ denote out-of-plane and in-plane total energies, respectively.

**Fabrication of devices and electrical measurements:** The $(Co_{0.15}Fe_{0.85})_5GeTe_2$ (CFGT) nanolayer flakes (with a thickness of 20-30 nm), were exfoliated and dry-transferred onto a CVD graphene channel on an $n^{++}Si/SiO_2$ (285 nm) substrate inside a N2 glovebox. CVD graphene channels were prepared by electron beam lithography (EBL) and oxygen plasma patterning. For the fabrication of spin valve devices, non-magnetic (Au/Ti) and magnetic contacts (Co/TiO$_2$) were prepared using multiple EBL processes and electron beam evaporation of metals. The Au/Ti contacts were first evaporated on CFGT flakes after a few seconds of Ar ion milling to clean the surface. After which, another round of EBL and Au/Ti evaporation was performed for reference electrodes in graphene. Lastly, the ferromagnetic contacts of Co (60 nm)/TiO$_2$(~1-2 nm) on graphene were prepared using a two-step deposition process. Specifically, 0.4 nm of Ti was deposited two times, followed by a 10 Torr O$_2$ oxidation for 10 minutes each, and then followed by 60 nm of Co deposition. The magnetic Co/TiO$_2$ contacts were designed with varying widths (400 - 500 nm) to serve as reference spin injector (detector), taking advantage of the well-defined magnetic properties of Co with in-plane easy axis controlled by strong shape anisotropy. The channel length and width of graphene in Dev 1 were ~ 4.5 μm and ~ 3 μm, respectively. The CFGT/Gr interface resistance was in the range of 1-3 kΩ, while the resistance of Co/TiO$_2$/Gr contacts were around 10-20 kΩ. For Dev 2, the graphene channel length and width were ~13.7 μm and ~3 μm, respectively. The CFGT/Gr interface resistance ranged from ~150 Ω to 200 Ω, while the resistance of Co/TiO$_2$/Gr contacts were around 7-25 kΩ.

The measurements were carried out at room temperature under vacuum conditions using magnetic field sweep and a sample rotation stage. The electronic measurement system is composed of a current source (Keithley 6221), a nanometer (Keithley 2182A), and a dual-channel source meter (Keithley 2612B).

**SQUID Measurements:** A Quantum Design superconducting quantum interference device (SQUID) magnetometer was used to measure the static magnetic properties of bulk CFGT crystals. The bulk crystal was attached to a Si substrate to properly align the magnetic field during hysteresis measurements in both in-plane (B∥ab) and out-of-plane (B∥c) configurations.

**Acknowledgments**

The authors acknowledge financial support from EU Graphene Flagship (Core 3, No. 881603), AoA Nano program at Chalmers, 2D TECH VINNOVA competence center (No. 2019-00068), Swedish Research Council VR project grants (Nos. 2021–04821 and 2021-0465), FLAG-ERA project 2DSOTECH (VR No. 2021-05925), Wallenberg Initiative Materials Science for Sustainability (WISE), We acknowledge the help of staff at the Quantum Device Physics and Nanofabrication laboratory in our department at Chalmers. B. S. acknowledges financial support from Swedish Research Council project grant 2022-04309. The computations were enabled in project SNIC 2022/3-30 by resources provided by the Swedish National Infrastructure for Computing (SNIC) at NSC and PDC, partially funded by the VR (No. 2018–05973), and in project NAISS 2023/5-226 & 2023/5-238 provided by the National Academic Infrastructure for Supercomputing in Sweden (NAISS) at UPPMAX, funded by the Swedish Research Council through grant agreement no. 2022-06725. B. S. and S. E. acknowledge allocation of supercomputing hours by EuroHPC resources in Karolina supercomputer in Czech Republic, and Lumi supercomputer in Finland.

**Data availability**

The data that support the findings of this study are available from the corresponding authors on a reasonable request.

**Author information**

**Affiliations**

Department of Microtechnology and Nanoscience, Chalmers University of Technology, SE-41296, Göteborg, Sweden
Department of Physics and Astronomy, Uppsala University, Box-516, SE-75120 Uppsala, Sweden
Department of Materials Science and Engineering, Uppsala University, Box 35, SE-75103 Uppsala, Sweden
Department of Physics, Indian Institute of Technology Ropar, Roopnagar 140001, Punjab, India
Graphene Center, Chalmers University of Technology, SE-41296, Göteborg, Sweden

**Contributions**

R.N., B.Z., and S.P.D. conceived the idea, designed the experiments, fabricated and characterized the devices, and wrote the manuscript. R.G. and P.S. performed SQUID measurements. S.E., M.D., and B.S. performed the calculations, and analyzed theoretical data. A.M.H., L.S., L.B., and A.K. supported in experiments. All authors participated in the interpretation of data, compilation of figures and writing of the manuscript. S.P.D. supervised the research project.

**Corresponding author**

Correspondence to Saroj P. Dash, saroj.dash@chalmers.se

**Competing interests**

The authors declare no competing interests.